\documentclass[aps,prd,preprint,showpacs,groupedaddress,nofootinbib]{revtex4}
\pdfoutput=1
\usepackage{amssymb,amsbsy,color}
\usepackage{latexsym}
\usepackage{graphics}
\usepackage{graphicx}
\usepackage{psfrag}

\newcommand{\be}{\begin{equation}}
\newcommand{\ee}{\end{equation}}
\newcommand{\bea}{\begin{eqnarray}}
\newcommand{\eea}{\end{eqnarray}}
\newcommand{\ba}{\begin{array}}
\newcommand{\ea}{\end{array}}

\newcommand{\pref}[1]{(\ref{#1})}

\newcommand{\mpit}{M_0^2}
\newcommand{\tct}{2c_2a^2}

\def\bqry{\begin{eqnarray}}
\def\eqry{\end{eqnarray}}

\def\lbl{\label}

\begin{document}

\preprint{UTHEP-565, HU-EP-08/24, SFB-CCP-08-44}
\title{
Pion scattering in  Wilson ChPT  
}
\author{$^{1,2}$Sinya Aoki, $^{3}$Oliver B\"ar and $^{3}$Benedikt Biedermann}

\affiliation{
$^1$Graduate School of Pure and Applied Sciences, University of Tsukuba,
Tsukuba 305-8571, Ibaraki Japan \\
$^2$Riken BNL Research Center, Brookhaven National Laboratory, Upton,
NY 11973, USA \\
$^3$Institute of Physics, Humboldt University Berlin, Newtonstrasse
15, 12489 Berlin, Germany
}

\date{\today}
%
\begin{abstract}
%
We compute the scattering amplitude for pion scattering in Wilson chiral perturbation theory for two degenerate quark flavors. We consider two different regimes where the quark mass $m$ is of order (i) $a^{}\Lambda_{\rm QCD}^2$ and (ii) $a^{2}\Lambda_{\rm QCD}^3$. 
Analytic expressions for the scattering lengths in all three isospin channels are given.
As a result of the O$(a^{2})$ terms the $I=0$ and $I=2$ scattering lengths do not vanish in the chiral limit. Moreover, additional chiral logarithms proportional to $a^2\ln M_{\pi}^2$ are present in the one-loop results  for regime (ii). These contributions significantly modify the familiar results from continuum chiral perturbation theory.

\end{abstract}
\pacs{11.15.Ha, 12.39.Fe, 12.38.Gc}
\maketitle

\section{Introduction}
\label{sect:intro}

The scattering of low-energy pions is an important process in QCD. Since the pions are the pseudo Goldstone bosons associated with the spontaneous breaking of chiral symmetry, one can use chiral perturbation theory (ChPT) to compute the scattering amplitude and the associated scattering lengths to  high precision. These theoretical predictions are in excellent agreement with experiment \cite{Leutwyler:2006qq,Colangelo:2007df}.

Since a few years ago Lattice QCD has started to  impact ChPT and the study of pion scattering. Increase in computer power and algorithmic advances nowadays allow unquenched simulations with light pion masses so that contact with ChPT can be made. Monitoring the quark mass dependence of the pion mass and the pion decay constant various lattice groups have obtained estimates for the low energy constants $\bar{l}_3$ and $\bar{l}_4$ that enter the one-loop ChPT results of these two observables \cite{Aubin:2004fs,DelDebbio:2006cn,Boucaud:2007uk}. These two constants play an important role for pion scattering, in particular $\bar{l}_3$, since it dominates the uncertainties in the isospin zero and isospin two scattering lengths \cite{Colangelo:2001df} (see also \cite{Colangelo:2007df}).

The scattering lengths can also be computed directly on the lattice, and two unquenched calculations for the $I=2$ scattering length $a_0^2$ have been performed so far. The CP-PACS collaboration simulated two dynamical Wilson quarks at three lattice spacings between 0.1 and  0.2 fm \cite{Yamazaki:2004qb}. The pion masses, however, were rather heavy, spanning the range 0.5 to 1.1 GeV. Thus, contact to ChPT is not expected. The NPLQCD collaboration \cite{Beane:2007xs,Beane:2005rj} performed a mixed action simulation with domain-wall valence quarks and staggered sea quarks.\footnote{The unquenched configurations were generated by the MILC collaboration \cite{Bernard:2001av} and are publicly available.} The four pion masses were in the range 290 to 490 MeV and the lattice spacing $a\approx 0.125$ fm.\footnote{The pion masses refer to the staggered Goldstone pion. The masses of the other pions are significantly heavier. The taste singlet pion, which is the heaviest, is about 450 MeV heavier than the Goldstone pion at this lattice spacing \cite{Aubin:2004fs}.} Since data at one lattice spacing only was generated and due to the use of different actions in the sea and valence sector mixed action ChPT \cite{Bar:2002nr,Bar:2005tu,Chen:2006wf} was used to extrapolate simultaneously to the continuum and chiral limit. The obtained result  for $a_0^2$ agrees very well with the experimentally measured one.

In this paper we study pion scattering in Wilson ChPT (WChPT) \cite{Sharpe:1998xm,Rupak:2002sm}, the low-energy effective theory for Lattice QCD with Wilson quarks. We compute the scattering amplitude  to one loop including the leading corrections due to the non-zero lattice spacing. 
From the scattering amplitude we derive expressions for the scattering lengths in all three isospin channels which can be used to analyze lattice data before the continuum limit has been taken.
We work in two different quark mass regimes: (i) $m \sim a\Lambda_{\rm QCD}^2$ and (ii) $m \sim a^{2}\Lambda_{\rm QCD}^3$. These are most likely applicable to todays and future simulations with light Wilson quarks. The $I=2$ scattering length in regime (i) has already been calculated before in Ref.\ \cite{Buchoff:2008ve}, and we agree with this result.

The lattice spacing is, just as the quark mass, a source of explicit chiral symmetry breaking. Therefore, the lattice artifacts modify the results from continuum ChPT. For example, the scattering lengths $a_0^I$ for $I=0,2$ do not vanish in the chiral limit but assume non-zero values. This is not unexpected \cite{Kawamoto:1981hw,Sharpe:1992pp,Gupta:1993rn}. However, the value in the chiral limit is of order $a^2$ and not of order $a$, as one might naively think based on the (broken) symmetries of the Wilson fermion action. 

In addition, at sufficiently small quark masses there appear additional chiral logarithms proportional to $a^2\ln M_{\pi}^2$ in the chiral expansion. These additional contributions may obscure the continuum chiral logarithms.  Consequently, fits of the continuum ChPT expressions to the lattice data may easily result in erroneous determinations of the Gasser-Leutwyler coefficients associated with pion scattering. 

\section{Pion scattering at tree level}
\label{sect:tree}

\subsection{Setup}
\label{ssect:setup}

The chiral effective Lagrangian of WChPT is expanded in powers of (small) pion momenta $p^{2}$, quark masses $m$ and the lattice spacing $a$. 
Based on the symmetries of the underlying Symanzik action \cite{Symanzik:1983dc,Symanzik:1983gh} the chiral Lagrangian including all terms of  O$(p^{4},p^{2}m,m^2,p^{2}a,ma)$ is given in Ref.\ \cite{Rupak:2002sm}. The O$(a^{2})$ contributions are constructed in Ref.\ \cite{Bar:2003mh} and, independently,  in Ref.\ \cite{Aoki:2003yv} for the two-flavor case.  

In the following we will restrict ourselves to $N_{f}=2$  with degenerate quark mass $m$. In this case the chiral Lagrangian (in Euclidean space-time) including the $p^{2}, m$ and $a^2$ terms is found to be \cite{Bar:2003mh,Aoki:2003yv}
\bqry
\lbl{L2}
{\cal L}_{2} & =& \frac{f^{2}}{4} \langle \partial_{\mu}\Sigma \partial_{\mu}\Sigma^{\dagger}\rangle  - \frac{f^{2}}{4} \hat{m} \langle \Sigma + \Sigma^{\dagger}\rangle - \frac{1}{2}(2W_6^\prime+W_8^\prime)\hat{a}^2\langle{\Sigma+\Sigma^\dagger}\rangle^2,
\eqry
where the angled brackets denote traces over the flavor indices. The field $\Sigma$ containing the pion fields is defined as usual, 
\bqry
\lbl{Sigma}
\Sigma(x) &=& \exp \left( \frac{2 i}{f} \pi(x)\right), \qquad \pi(x)\,=\, \pi^{a}(x) \frac{\sigma^a}{2},
\eqry
with the standard Pauli matrices $\sigma^a$. The quark mass and the lattice spacing enter through the combinations
\bqry
\lbl{qmass}
\hat{m} & = & 2Bm\,,\qquad \hat{a} \, =\, 2W_0 a\,.
\eqry
The coefficients $B$ and $f$ are the familiar leading order (LO) low-energy coefficients (LECs) from continuum ChPT \cite{Gasser:1983yg,Gasser:1984gg}, while $W_0, W_6^{\prime}, W_8^{\prime}$ are LECs associated with the non-zero lattice spacing artifacts \cite{Rupak:2002sm,Bar:2003mh}.

Note that the mass parameter $m$ in eq.\ \pref{qmass} is the so-called {\em shifted mass} \cite{Sharpe:1998xm}. Besides the dominant additive mass renormalization proportional to 1/$a$ it also contains the correction of O($a$). Consequently, there is no term linear in the lattice spacing present in the chiral Lagrangian in eq.\ \pref{L2}.

Note also that we keep the O$(a^2)$ correction in ${\cal L}_2$ and promote it to a LO term in the chiral expansion. This is justified (and necessary) for small enough quark masses such that the O($m$) and the O$(a^2)$ terms are of the same order of magnitude, i.e.\ for $m\sim a^2\Lambda_{\rm QCD}^3$. 
We call this scenario the {\em large cut-off effects} (LCE) regime, in contrast to the so-called {\em generic small quark mass} (GSM) regime \cite{Sharpe:2004ps,Sharpe:2004ny}, which assumes $m\sim a\Lambda_{\rm QCD}^2$. Nevertheless, even though we almost exclusively work in the LCE regime we will be able to obtain the corresponding results for the GSM regime as well by appropriately expanding our  final results (see section \ref{ssect:GSM}). 

It will be useful to introduce \cite{Aoki:2004ta}
\bqry
\lbl{c2}
c_2 & = & - 32(2W_6^\prime+W_8^\prime)\frac{W_0^2}{f^2}
\eqry
for the combination of LECs in front of the O($a^2$) term in the chiral Lagrangian. The sign of $c_2$ determines the phase diagram of the theory \cite{Sharpe:1998xm}.\footnote{Note that our definition for $c_2$ differs by a factor $f^2a^2$ from the one in Ref.\ \cite{Sharpe:1998xm}.} For $c_2>0$ there exists a second-order phase transition separating a phase where parity and flavor are spontaneously broken \cite{Aoki:1983qi}. The charged pions are massless in this phase due to the spontaneous breaking of the flavor symmetry. Outside this phase the pion mass is given by
\bqry
\lbl{Mpitree}
M_0^2 & =& 2Bm - 2c_2a^2\,,
\eqry
and it vanishes at $m = c_2a^2/B$. For even smaller values of $m$ the charged pions remain massless, while the neutral pion becomes massive again \cite{Sharpe:1998xm}.

Negative values of $c_2$, on the other hand, imply a first order phase transition with a minimal non-vanishing pion mass. All three pions are massive for all quark masses, and the pion mass assumes its minimal value at $m=0$, resulting in 
\bqry
\lbl{Mpimin}
M_{0,{\rm min}}^2 & =& 2|c_2|a^2\,.
\eqry
The magnitude and the sign of $c_2$ depend on the details of the underlying lattice theory, i.e.\ what particular lattice action has been chosen. However, it is not a simple task to measure $c_2$ numerically. Adding a twisted mass term \cite{Frezzotti:2000nk} to the theory the pion mass splitting between the neutral and the charged pion is equal to $2c_2a^2$ at LO in the chiral expansion \cite{Scorzato:2004da,Aoki:2004ta}. This has been exploited by the ETM collaboration \cite{Michael:2007vn,Urbach:2007rt} and an estimate for $c_2$ has been obtained. Within errors $c_2$ is negative if the standard Wilson fermion action and the tree-level Symanzik improved gauge action is used. However, the statistical uncertainties for $c_2$  were rather large due to the presence of disconnected diagrams in the calculation of the neutral pion mass.  

\subsection{Scattering amplitude and scattering lengths}
\label{ssect:treeamplitude}
We are interested in the two-pion scattering process
\bqry
\lbl{pipiprocess}
\pi^{\alpha}(p) +  \pi^{\beta}(k) \longrightarrow \pi^{\gamma}(p') + \pi^{\delta}(k')\,.
\eqry
This process is described by the scattering amplitude $A$. It is convenient to use the three Mandelstam variables as arguments for it, $A=A(s,t,u)$. The scattering amplitude is  straightforwardly calculated as the residue of the four pion pole in the four-point function. Starting from eq.\ \pref{L2}  we obtain the tree-level result\footnote{Recall that the Lagrangian in Eq.\ \pref{L2} is given in Euclidean space-time, so one has to Wick-rotate back to Minkowski space in order to get the physical scattering amplitude.}
\bqry
\lbl{amplitudetree}
A(s,t,u) & = & \frac{1}{f^2}(s - \mpit - \tct)\,.
\eqry
Setting the lattice spacing to zero we recover, as expected, the familiar result of continuum ChPT \cite{Gasser:1983yg}.

Having calculated the scattering amplitude we perform the standard partial wave expansion and obtain the scattering lengths \cite{Gasser:1983yg}. Most interesting are the scattering lengths $a_0^I$ for definite isospin $I=0$ and $I=2$:
\bqry
\lbl{treescatlen}
a_0^0 & =&\phantom{-} \frac{7}{32\pi f^2} \left(\mpit - \frac{5}{7}\, \tct\right)\,,\lbl{treescatlen0}\\
a_0^2 & = &  - \frac{1}{16\pi f^2} \Bigg(\mpit + \tct \Bigg)\,.\lbl{treescatlen2}
\eqry
Again, for $a=0$ we recover the tree-level continuum results, first obtained by Weinberg \cite{Weinberg:1966kf}. For a non-zero lattice spacing, however, the continuum results are modified in such a way that the scattering lengths no longer vanish in the chiral limit. Instead, they assume non-zero values of  O($a^2$).  
In other words, the ratio $a_0^I/M_{\pi}^2$ is no longer a constant but has the functional form
\bqry
\lbl{treeadivmpi}
\frac{a_0^I}{M_{\pi}^2} & =& \frac{A_{00}^I}{M_{\pi}^2} + A^I_{10}\,,
\eqry
with $A^I_{10}$ being a constant and $A^I_{00}$ being of order $a^2$. Hence, the ratio $a_0^I/M_{\pi}^2$ diverges in the chiral limit. 
This divergence has been anticipated first by Kawamoto and Smit \cite{Kawamoto:1981hw}. However, note that the coefficient $A^I_{00}$ is of order $a^2$ rather than of order $a$. This holds even for  unimproved Wilson fermions, in contrast to earlier expectations \cite{Sharpe:1992pp,Gupta:1993rn}.

\begin{figure*}[t]
\begin{center}
 \includegraphics[scale=0.85]{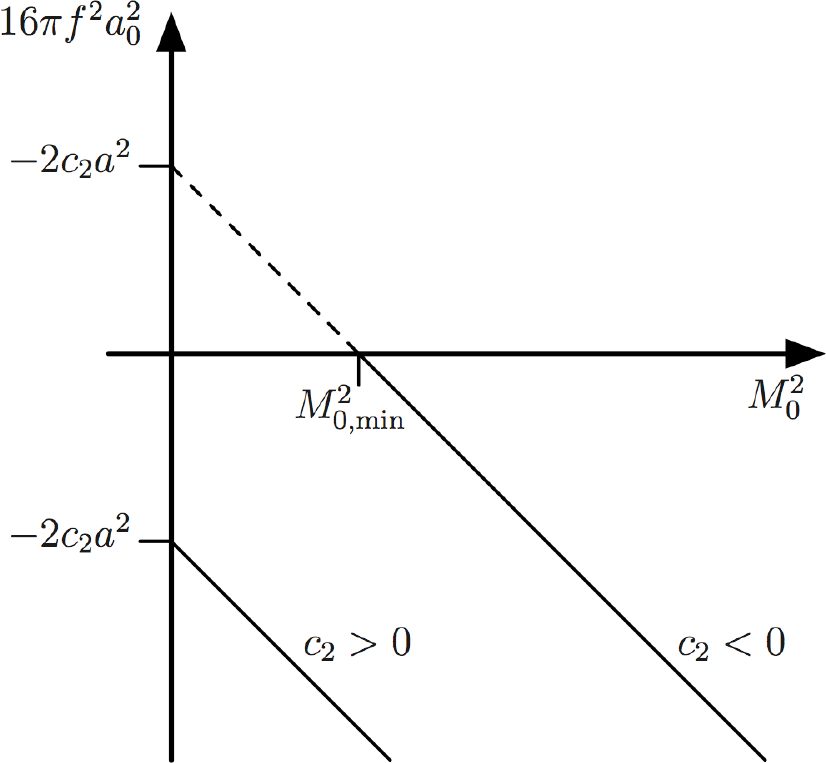}
 ~\hfill~
\includegraphics[scale=0.85]{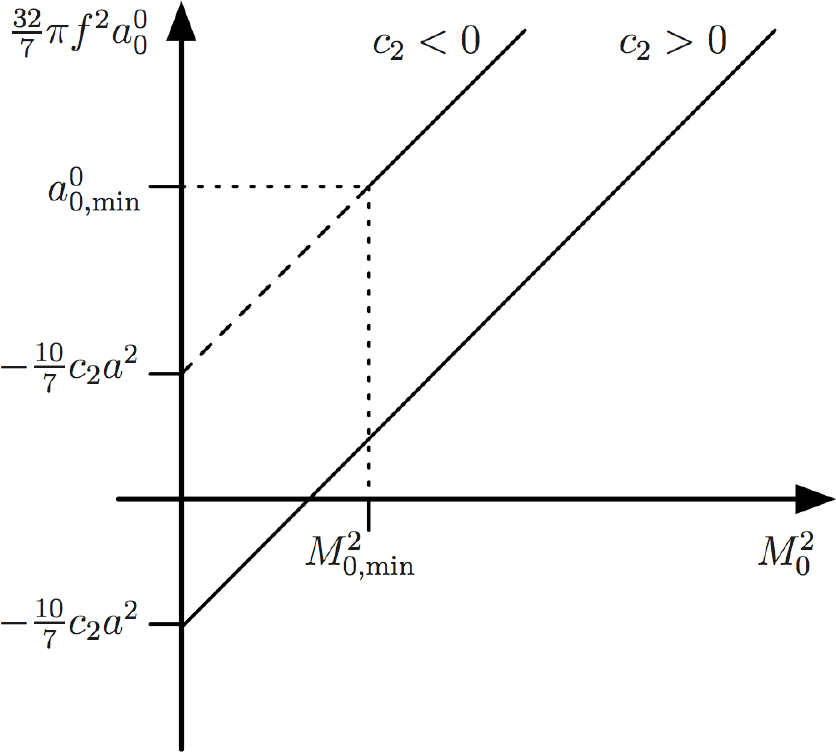}
 \end{center}
\caption{Sketch of the scattering lengths as a function of the pion mass at non-zero lattice spacing.
The left panel shows $16\pi f^2a_0^2$ as a function of the tree level pion mass $M_0^2$. For $c_2<0$ the pion mass cannot be smaller than $M^2_{0,{\rm min}}$ in eq.\ \pref{Mpimin}. Nevertheless, extrapolating to the massless point the scattering length assumes the value $-2c_2a^2$, as indicated by the dashed line. For $c_2>0$ the pion mass can be taken zero. At this mass the scattering length also assumes the value $-2c_2a^2$, now with the opposite sign.
The right panel shows the analogous sketch for the $I=0$ scattering length.
 }
\label{fig:1}      
\end{figure*}

On the other hand, the divergence in the chiral limit
will only be present if $c_2>0$, because only in this case can the pion indeed become massless. For the opposite sign the pion mass cannot be smaller than the minimal value quoted in eq.\ \pref{Mpimin}, resulting in the following minimal values for the scattering lengths:
\bqry
\lbl{a0min}
a_{0,{\rm min}}^0 & =& \frac{12}{32\pi f^2} 2|c_2|a^2\,,\qquad
a_{0,{\rm min}}^2 \, = \, 0\,.
\eqry
Figure \ref{fig:1} sketches the pion mass dependence of the scattering lengths for the two possible signs of $c_2$. It seems feasible that measurements of  the scattering lengths will allow more precise determinations of  $c_2$. Extrapolating the data for $a_0^2$ to the chiral limit  one may directly read off  $c_2$ as the value at vanishing pion mass, even for the $c_2<0$ case. A practical advantage is that the calculation of $a_0^2$ does not involve disconnected diagrams which introduce large statistical uncertainties.

The $I=1$ channel is somewhat special. The scattering amplitude for this isospin channel is given by  $A(t,s,u)-A(u,t,s)$. The $c_2$ contribution drops out in this difference and the scattering length $a_1^1$ is is given by
\bqry
\lbl{treescatlen1}
a_1^1 & = &   \frac{1}{24\pi f^2} \mpit\,.
\eqry
This is exactly the tree level result of continuum ChPT \cite{Gasser:1983yg} and it suggests that the scaling violations in $a_1^1$ are very small.\footnote{Note that our definition of the scattering length $a_1^1$ differs by a factor of  $\mpit$ from the one in Ref.\ \cite{Gasser:1983yg}.} 
 
\section{Pion scattering at one loop}
\label{sect:loop}

\subsection{Power counting and higher order terms}
\label{ssect:pc}

\begin{figure*}[t]
\begin{center}
\includegraphics[scale=0.9]{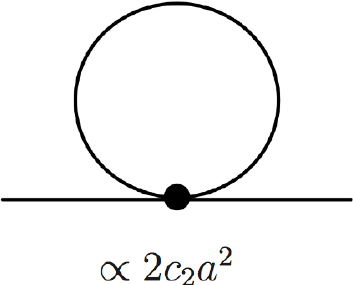}
 \end{center}
\caption{Additional one-loop diagramm contributing to the self energy.
The vertex is proportional $c_2a^2$, leading to a chiral logarithm $ c_2a^2\ln M_{0}^2$.
 }
\label{fig:2}      
\end{figure*}

Having promoted the O($a^2$) correction to the LO Lagrangian, it will contribute to the one-loop results of various quantities in a non-trivial way. Expanding the O($a^2$) term in terms of pion fields we obtain vertices proportional to $2c_2a^2$. As part of one-loop diagrams these lead to non-analytic corrections proportional to $2c_2a^2\ln M^2_{0}$, for example in the expression of the pion mass (see figure \ref{fig:2}). 

The presence of additional chiral logarithms of order $a^2\ln M^2_{0}$ is well known in staggered ChPT (SChPT) \cite{Bernard:2001yj,Aubin:2003mg,Aubin:2003uc}, and has been pointed out first in Ref.\ \cite{Aoki:2003yv} for WChPT. They are considered to be one reason why the chiral logarithms known from continuum ChPT are not reproduced in the lattice data: The additional chiral logs obscure the non-analytic quark mass dependence due to the continuum chiral logs, and the naively expected behaviour is lost.

The power counting in WChPT is slightly non-trivial if we take the O($a^2$) term at LO. The LO Lagrangian consists of the terms of O($p^2,m,a^2$). In order to renormalize the divergencies of the loop diagrams we need higher order counterterms in the chiral Lagrangian. These are, besides the standard ones in ${\cal L}_4$ of continuum ChPT \cite{Gasser:1983yg,Gasser:1984gg}, the terms of order $p^2a^2, ma^2,a^4$. However, terms of order $p^2a, ma,a^3$ are also present and formally of lower order. Hence these should also be included, even though their LECs do not get renormalized at one loop. To conclude, one-loop calculations in the LCE regime require the following terms in the chiral Lagrangian:\\
\underline{LCE regime:}
\vspace{-0.8cm}
\bqry
\lbl{PcountingLCE}
\begin{array}{rcl}
{\rm LO:}& \quad & p^2,\,m,\,a^2\\
{\rm NLO:}& \quad & p^2a,\,ma,\,a^3\\
{\rm NNLO:}& \quad & p^4,\,p^2m,\,m^2,\,p^2a^2,\, ma^2,\,a^4
\end{array}
\eqry
The standard NLO terms of continuum ChPT appear here at NNLO, a consequence of the fact that $m\sim a^2\Lambda_{\rm QCD}^3$ in the LCE regime.\footnote{The terms of O($p^2a,ma,a^3)$ are not present in SChPT due to the axial U(1) symmetry. Consequently, the terms of the third row in \pref{PcountingLCE} are the NLO terms.} In the GSM regime, however, we recover the standard ordering. Since $m\sim a\Lambda_{\rm QCD}^2 > a^2\Lambda_{\rm QCD}^3$ in this regime, a one-loop calculation requires the following terms:\\
\underline{GSM regime:}
\vspace{-0.8cm}
\bqry
\lbl{PcountingGSM}
\begin{array}{rcl}
{\rm LO:}& \quad & p^2,\,m \\
{\rm NLO:}& \quad & p^4,\,p^2m,\,m^2,\,p^2a,\, ma,\,a^2
\end{array}
\eqry  
Note that all terms in \pref{PcountingGSM} are also present in \pref{PcountingLCE}, even though reshuffled. This already indicates, that a result obtained in the LCE regime yields the corresponding result in the GSM regime if appropriately expanded.

Most of the necessary terms in \pref{PcountingLCE} have already been constructed. So far unknown are the contributions of order $p^2a^2, ma^2,a^3,a^4$. Performing a standard spurion analysis with the spurion fields in Ref.\ \cite{Bar:2003mh} it is straightforward to construct these missing terms, at least the ones we need as counterterms for the observables we are interested in here, the pion mass and the scattering amplitude. Some details concerning the construction of these terms are collected in appendix \ref{appA}.

Finally, if the underlying lattice theory is non-perturbatively O($a$) improved according to the Symanzik improvement program \cite{Symanzik:1983dc,Symanzik:1983gh}, the terms of order $p^2a,ma$ are absent in \pref{PcountingLCE} and \pref{PcountingGSM}. In the following we will always keep these contributions in our calculation; the O($a$) improved result is simply obtained by dropping the appropriate terms linear in the lattice spacing.

\subsection{Pion mass to one loop}
\label{ssect:pionmassoneloop}

The modification due to the additional chiral logarithms is illustrated best in the one-loop result for the pion mass, which we will also need in the next section. In terms of the tree-level pion mass $M_0^2$, cf.\ Eq.\ \pref{Mpitree}, we find
\bqry
\lbl{pionmassoneloop}
 M_{\pi}^2&=&M^2_0\left[1+\frac{1}{32\pi^2}\frac{M^2_0}{f^2}\ln\left(\frac{M^2_0}{\Lambda_3^2}\right)+\frac{5}{32\pi^2}\frac{2c_2a^2}{f^2}\ln\left(\frac{M^2_0}{\Xi_3^2}\right) +k_1\frac{W_0a}{f^2}\right]\nonumber \\
 &&\,+\,k_3\frac{2c_2W_0a^3}{f^2}+k_4\frac{(2c_2a^2)^2}{f^2}\,.
\eqry
The coefficients $\Lambda_3^2,\Xi_3^2$ and $k_1,k_3,k_4$ are (combinations of) unknown LECs. 
$\Lambda_3^2$ is the familiar scale independent LEC present in the continuum result \cite{Gasser:1983yg}, and in complete analogy we introduced the new LEC $\Xi_3$. Note that we have chosen the coefficients $k_i$ to be dimensionless.\footnote{Details about how the LECs in \pref{pionmassoneloop}   are related to the ones in the chiral Lagrangian can be found in \cite{DABied}.} In the O($a$) improved theory the $k_1$ term is absent.

A few comments concerning \pref{pionmassoneloop} are in order. First, setting the lattice spacing to zero we recover the familiar result from continuum ChPT \cite{Gasser:1983yg}, as expected. This result gets modified at non-zero lattice spacing. In particular, there exists the anticipated chiral logarithm proportional to $c_2a^2\ln M^2_0$. Note that the coefficient in front of it is ten times larger than the coefficient in front of the continuum chiral log proportional to $M_0^2\ln M^2_0$. Hence, even small O($a^2$) contributions may dilute the continuum chiral logarithm completely. This is better seen if we rewrite the square bracket in \pref{pionmassoneloop} as 
\bqry
\lbl{squarebracket}
 \left[1+\frac{1}{32\pi^2}\{M^2_0 + 10 c_2a^2\}\ln\left(\frac{M^2_0}{\Lambda_3^2}\right)+\frac{10}{32\pi^2f^2}c_2a^2\ln\left(\frac{\Lambda_3^2}{\Xi_3^2}\right) +k_1\frac{W_0a}{f^2}\right]\,,
 \eqry
so that the quark mass dependence comes entirely from the $\ln M_0^2/\Lambda_3^2$ term.
Negative values of $c_2$ can render the factor $M^2_0 + 10 c_2a^2$ exceptionally small such that the chiral logarithm is effectively not active. 

This scenario is not as unlikely as one may think. We remark that the ETM collaboration has found a negative value for $c_2$ in their twisted mass simulations, and the calculation of the pion mass splitting provides a rough estimate for $-2c_2a^2 =M^2_{\pi^{\pm}} - M^2_{\pi^{0}}$. The data  \cite{Urbach:2007rt} for $a \approx 0.086$ fm and $M_{\pi^{\pm}}\approx 300$ MeV results in $-2c_2a^2 \approx (185$ MeV$)^2$. Such a value completely suppresses the chiral log for pion masses around 400 MeV, a value not unusual in lattice simulations performed these days.

Finally, note that the pion mass \pref{pionmassoneloop} does not vanish in the limit $M_0^2=0$ because of the corrections proportional to $k_3,k_4$. However, these are contributions of O($a^3,a^4$) to the additive mass renormalization (the critical quark mass) and can be absorbed by an appropriate finite renormalization. To be specific, define $\tilde{m}$ and $\tilde{M}_0^2$ by
\bqry
\lbl{massrenorm}
\tilde{M}_0^2 \,=\, 2B\tilde{m} & \equiv & 2Bm -2c_2a^2 + k_3 \frac{2c_2W_0a^3}{f^2} + \left(k_4\frac{(2c_2a^2)^2}{f^2} - k_1 \frac{W_0a}{f^2}k_3\frac{2c_2W_0a^3}{f^2}\right)\,,
\eqry
such that the quark masses $\tilde{m}$ and $m$ differ by order $a^2$ and higher. In terms of $\tilde{m}$ the result \pref{pionmassoneloop} reads (up to the order we are working here)
\bqry
\lbl{pionmassoneloop2}
 M_{\pi}^2&=&\tilde{M}^2_0\left[1+\frac{1}{32\pi^2}\frac{\tilde{M}^2_0}{f^2}\ln\left(\frac{\tilde{M}^2_0}{\Lambda_3^2}\right)+\frac{5}{32\pi^2}\frac{2c_2a^2}{f^2}\ln\left(\frac{\tilde{M}^2_0}{\Xi_3^2}\right) +k_1\frac{W_0a}{f^2}\right]\,.
\eqry
The $k_3,k_4$ contribution no longer appear explicitly but are absorbed in the definition of the quark mass $\tilde{m}$. With this parametrization the pion mass vanishes in the chiral limit for $\tilde{m}=0$. Therefore, $\tilde{m}$ is proportional to the subtracted bare lattice quark mass $(m_0 -m_{\rm cr})$ if the critical quark mass $m_{\rm cr}$ is defined by a vanishing pion mass for $m_0 = m_{\rm cr}$. In other words, doing the definition \pref{massrenorm} we have appropriately matched the chiral effective theory to the underlying lattice theory for this particular renormalization condition. Obviously, the matching differs for other definitions of the lattice quark mass, for example through the PCAC relation.

In the following we are not interested in the quark mass dependence of scattering observables, but rather in the dependence on the pion mass. For this the parametrization in terms of $m$ is sufficient, since at the end we will replace $m$ by $M_{\pi}^2$ using Eq.\ \pref{pionmassoneloop}.

\subsection{Scattering at one loop}
\label{ssect:scatteringoneloop}

\begin{figure*}[t]
\begin{center}
\includegraphics[scale=0.65]{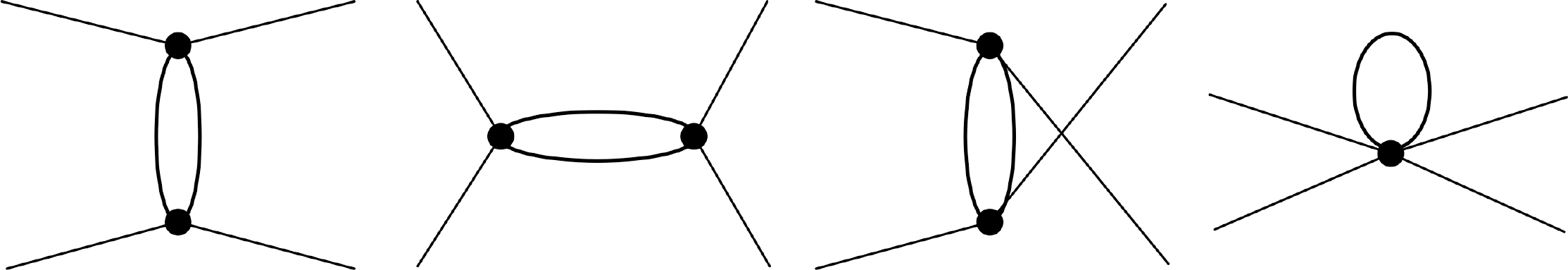}
 \end{center}
\caption{One-loop diagramms contributing to the four-point function.
The diagrams with one or two vertices stemming from the O($a^2$) term in the chiral Lagrangian will give rise to chiral logs proportional $ c_2a^2M_{0}^2\ln M_{0}^2$ and $(c_2a^2)^2\ln M_{0}^2$.
 }
\label{fig:3}      
\end{figure*}

At one loop there contribute the four diagrams in figure \ref{fig:3} to the four-point function. The vertices in these diagrams can be either a vertex also present in continuum ChPT or the vertex proportional to $c_2a^2$ stemming from the O($a^2$) term in Eq.\ \pref{L2}. Hence, we expect additional chiral logarithms $c_2a^2M_0^2\ln M_0^2 $  and $(c_2a^2)^2\ln M_0^2$ in the scattering amplitude and the scattering lengths. 

The one-loop result for the scattering amplitude can be written as
\bqry
\lbl{scatamponeloop}
A(s,t,u)& =&\frac{1}{f^2}(s-\mpit-2c_2a^2)+B(s,t,u)+C(s,t,u)\,,
\eqry
where the functions $B$ and $C$ contain the one-loop corrections. The general structure of these functions is
\bqry
\lbl{BCfunction}
B(s,t,u) & =& B_{\rm cont}(s,t,u) + 2c_2a^2 B_{a^2}(s,t,u)\,,\lbl{Bfunction}\\
 C(s,t,u)& =& C_{\rm cont}(s,t,u) + 2c_2a^2 C_{a^2}(s,t,u)\,.\lbl{Cfunction}
\eqry
The first parts $B_{\rm cont},C_{\rm cont}$ are the contributions from continuum ChPT and can be found in Refs.\ \cite{Gasser:1983kx,Gasser:1983yg}. The contributions $B_{a^2},C_{a^2}$ come from the O($a^2$) correction. Since these expressions are cumbersome and not very illuminating we present them in appendix \ref{appB}.
Here we just mention that these functions are finite in the limit $a\rightarrow 0$, so in the continuum limit we recover the continuum result for the scattering amplitude.

With the amplitude at hand we can compute the scattering lengths as before. As a short-hand notation we introduce 
\bqry
\bar{l}_i(M_0^2) & = &- \ln \left(\frac{M_0^2}{\Lambda_i^2}\right)\,,\qquad i\,=\,1,2,3\lbl{Deflogl}\\
\bar{\xi}_i(M_0^2)&=& - \ln \left(\frac{M_0^2}{\Xi_i^2}\right)\,,\qquad i\,=\,3,\ldots,6 \lbl{Deflogxi}
\eqry
for the chiral logarithms in the following expressions. The $\bar{l}_i$ are the standard scale invariant LECs of continuum ChPT \cite{Gasser:1983yg}, and the parameters $\bar{\xi}_i $ are defined in complete analogy to them. However, we have made the dependence on $M_0^2$ explicit. This seems appropriate since the pion mass is varied in lattice simulations and  the $\bar{l}_i,\bar{\xi}_i$ are indeed functions of the pion mass.

With these definitions the one-loop results for the isospin zero and isospin two scattering lengths read
\bqry
\lbl{resulta0}
 a_0^0&=&\phantom{-}\frac{7}{32\pi}\frac{M_0^2}{f^2}\left(1+\frac{5}{84\pi^2}\frac{M_0^2}{f^2}\left\{\bar{l}_1(M_0^2)+2\bar{l}_2(M_0^2)+\frac{21}{8}\right\}\right.\nonumber\\
 &&\phantom{\frac{7}{32\pi}\frac{M_0^2}{f^2}\bigg(1+}\left.-\frac{5}{84\pi^2}\frac{2c_2a^2}{f^2}\left\{9\bar{\xi}_4(M_0^2)-\frac{33}{8}\bar{\xi}_5(M_0^2)+\frac{15}{4}\right\}\right)\nonumber\\
 &&-\frac{5}{32\pi}\frac{2c_2a^2}{f^2}\left(1-\frac{W_0a}{f^2}k_5-\frac{2c_2a^2}{32\pi^2f^2}\left\{11\bar{\xi}_6(M_0^2)-1\right\} \right)\nonumber\\
 &&+\frac{3}{8\pi f^2}(M_{\pi}^2 - M_0^2)\,,\\[2ex]
 a_0^2&=&-\frac{M_0^2}{16\pi
 f^2}\left(1-\frac{M_0^2}{12\pi^2f^2}\left\{\bar{l}_1(M_0^2)+2\bar{l}_2(M_0^2)+\frac{3}{8}\right\}-\frac{2c_2a^2}{32\pi^2f^2}\left\{11\bar{\xi}_5(M_0^2)+2\right\}\right)\nonumber\\
 &&-\frac{2c_2a^2}{16\pi
 f^2}\left(1-\frac{W_0a}{f^2}k_5-\frac{2c_2a^2}{32\pi^2f^2}\left\{11\bar{\xi}_6(M_0^2)-7\right\}\right)\,.\lbl{resulta2}
\eqry
Here $k_5$ is another LEC associated with the O($a^3$) terms in the chiral Lagrangian. As anticipated, we find additional chiral logarithms $a^2M_0^2\ln M_0^2 $ (the terms involving $\bar{\xi}_4,\bar{\xi}_5$) and $a^4\ln M_0^2$ (the $\bar{\xi}_6$ term).

The results \pref{resulta0} and \pref{resulta2} are given as a function of the quark mass $m$ via the tree-level pion mass $M_0^2$, except for the last line in $a_0^0$, where the one-loop expression $M_{\pi}^2$ enters.\footnote{The definition of $a_0^0$ involves the physical pion mass when one goes on-shell and uses $s+t+u=4M_{\pi}^2$.} 
As already mentioned at the end of section \ref{ssect:pionmassoneloop}, these results have to be properly matched to the lattice theory. Depending on the particular renormalization condition for the quark mass the final results differ by terms of order $a$. For this reason it is advantageous to express the scattering lengths as a function of the pion mass and not of the quark mass. Using the one-loop result \pref{pionmassoneloop} in order to replace $M_0^2$ by $M_{\pi}^2$ we obtain the scattering lengths as a function of $M_{\pi}^2$:
\bqry
\lbl{resulta0two}
 a_0^0&=&\phantom{-}\frac{7}{32\pi}\frac{M_{\pi}^2}{f^2}\left(1+\frac{5}{84\pi^2}\frac{M_{\pi}^2}{f^2}\left\{\bar{l}_1(M_{\pi}^2)+2\bar{l}_2(M_{\pi}^2)-\frac{3}{8}\bar{l}_3(M_{\pi}^2)+\frac{21}{8}\right\}\right.\nonumber\\
 &&\phantom{\frac{7}{32\pi}\frac{M_{\pi}^2}{f^2}\bigg(1+}\left.+\frac{5}{7}k_1\frac{W_0a}{f^2}-\frac{5}{84\pi^2}\frac{2c_2a^2}{f^2}\left\{\frac{15}{8}\bar{\xi}_3(M_{\pi}^2)+ 9\bar{\xi}_4(M_{\pi}^2)-\frac{33}{8}\bar{\xi}_5(M_{\pi}^2)+\frac{15}{4}\right\}\right)\nonumber\\
 &&-\frac{5}{32\pi}\frac{2c_2a^2}{f^2}\left(1-\left\{k_3+k_5\right\}\frac{W_0a}{f^2} -k_4\frac{2c_2a^2}{f^2}-\frac{2c_2a^2}{32\pi^2f^2}\left\{11\bar{\xi}_6(M_{\pi}^2)-1\right\} \right)\,,\\[2ex]
 a_0^2&=&-\frac{M_{\pi}^2}{16\pi
 f^2}\left(1-\frac{M_{\pi}^2}{12\pi^2f^2}\left\{\bar{l}_1(M_{\pi}^2)+2\bar{l}_2(M_{\pi}^2)-\frac{3}{8}\bar{l}_3(M_{\pi}^2)+\frac{3}{8}\right\}\right.\nonumber\\
 && \left.\phantom{\frac{M_{\pi}^2}{16\pi
 f^2}\big(1+}-k_1\frac{W_0a}{f^2} -\frac{2c_2a^2}{32\pi^2f^2}\left\{-5\bar{\xi}_3(M_{\pi}^2)+11\bar{\xi}_5(M_{\pi}^2)+2\right\}\right)\nonumber\\
 &&-\frac{2c_2a^2}{16\pi
 f^2}\left(1-\left\{k_3+k_5\right\}\frac{W_0a}{f^2}-k_4\frac{2c_2a^2}{f^2}-\frac{2c_2a^2}{32\pi^2f^2}\left\{11\bar{\xi}_6(M_{\pi}^2)-7\right\}\right)\,.\lbl{resulta2two}
\eqry
Here the coefficients $\bar{l}_i$ and $\bar{\xi}_i$ are as in \pref{Deflogl}, \pref{Deflogxi} but with the tree-level pion mass replaced by $M_{\pi}^2$. 

Note that we have still expressed the scattering lengths as a function of $f$, the decay constant in the chiral and continuum limit, and not in terms of $f_{\pi}$. The reason is that $f_{\pi}$ has not been computed yet to one loop in the LCE regime.\footnote{The calculation of the decay constant poses additional complications: The axial vector current in the effective theory needs to be constructed to the appropriate order and the proper renormalization condition has to be taken into account \cite{Aoki:2007es}.}  This is not an impediment to our results, but it means that $f$ is a free fit parameter when our results are used to fit them to numerical lattice data.\footnote{Note that the replacement of $f$ by $f_{\pi}$ does not reduce the number of unknown LECs since the latter contains new LECs that are not present in the expressions paramterized by $f$ \cite{Aoki:2007es}.} We briefly come back to this in section \ref{ssect:Disc}.

Our final results for the scattering lengths differ significantly from the corresponding ones in continuum ChPT. The scattering lengths do not vanish in the chiral limit, but rather diverge as $a^4\ln M_{\pi}^2$. 
However, if $c_2 <0$ there is no divergence since the pion mass cannot become smaller than the minimal value in eq.\ \pref{Mpimin}. For the opposite sign massless pions are in principle possible, but for small pion masses the chiral expansion eventually breaks down when $c_2a^2\ln M_{\pi}^2$ becomes of order unity. In that case higher order terms leading to powers $(c_2a^2\ln M_{\pi}^2)^n, n=2,3,\ldots$ become relevant too and a summation of all these terms is necessary. This can presumably be done along the lines in Ref.\ \cite{Aoki:2003yv}, where a resummed pion mass formula has been derived. 
Here, however, we assume that the pion masses are heavy enough such that the chiral logarithm $c_2a^2\ln M_{\pi}^2$ is a reasonably small correction to the leading order contribution. This is most likely the relevant case for actual numerical simulations.

In any case, the pion mass dependence of the scattering lengths on the lattice can be very different in contrast to what one may expect from continuum ChPT. Consequently, attempts to fit lattice data using expressions from continuum ChPT may easily fail.

\subsection{Results for the GSM regime}
\label{ssect:GSM}

In this section we summarize the results for the GSM regime where $m\sim a\Lambda_{\rm QCD}^2>a^2\Lambda_{\rm QCD}^3$. These are easily obtained from the expressions in the last section by expanding the logarithms according to $\ln(M_0^2) \approx \ln(2Bm)  -2c_2a^2/2Bm$ and dropping consistently all higher order terms. 

For example, the one-loop expression \pref{pionmassoneloop} for the pion mass reduces to 
\bqry
\lbl{pionmassGSM}
 M_{\pi}^2&=&2Bm\left[1+\frac{2Bm}{32\pi^2f^2}\ln\left(\frac{2Bm}{\Lambda_3^2}\right)+k_1\frac{W_0a}{f^2}\right]-2c_2 a^2\,.
 \eqry
This result is easily understood. The LO Lagrangian consists only of the kinetic and the mass term.
Hence, the tree-level pion mass is equal to $2Bm$ and the chiral logarithm in the one-loop result is just the one from continuum ChPT. The O($ma,a^2)$ terms enter at NLO and give analytic corrections only. 

Similarly we obtain the results for the scattering lengths:
\bqry
\lbl{a0GSM}
 a_0^0&=&\phantom{-}\frac{7}{32\pi}\frac{M_{\pi}^2}{f^2}\left(1+\frac{5}{84\pi^2}\frac{M_{\pi}^2}{f^2}\left\{\bar{l}_1(M_{\pi}^2)+2\bar{l}_2(M_{\pi}^2)-\frac{3}{8}\bar{l}_3(M_{\pi}^2)+\frac{21}{8}\right\}
 +\frac{5}{7}k_1\frac{W_0a}{f^2}\right)\nonumber\\
 &&-\frac{5}{32\pi}\frac{2c_2a^2}{f^2}\\[2ex]
  a_0^2&=&-\frac{M_{\pi}^2}{16\pi
 f^2}\left(1-\frac{M_{\pi}^2}{12\pi^2f^2}\left\{\bar{l}_1(M_{\pi}^2)+2\bar{l}_2(M_{\pi}^2)-\frac{3}{8}\bar{l}_3(M_{\pi}^2)+\frac{3}{8}\right\} - k_1\frac{W_0a}{f^2}\right)\nonumber\\
 &&-\frac{2c_2a^2}{16\pi
 f^2}\lbl{a2GSM}
\eqry 
As expected, the results reduce to the continuum one-loop result plus analytic corrections of order $M_{\pi}^2a$ and $a^2$. 

\subsection{Practical remarks}
\label{ssect:Disc}

An important application of WChPT is to provide formulae that can be used to fit lattice data and thereby allow the chiral extrapolation to the physical pion mass. As a by-product one also obtains estimates for the Gasser-Leutwyler coefficients involved in these formulae. The results for the scattering lengths of the previous sections serve exactly this purpose, but the way we have written them is not practical.
In order to compare and cross check our results with continuum ChPT we have parametrized the chiral logarithms in terms of $\bar{l}_i$ and  $\bar{\xi}_i$, defined in eqs.\ \pref{Deflogl} and \pref{Deflogxi}. Introducing the scale independent coefficients  $\bar{l}_i$ is useful in continuum ChPT, since the pion mass is constant in nature; consequently, the coefficients $\bar{l}_i$ are constants too. In lattice simulations we are free to choose the quark and pion mass, and it is this freedom that eventually enables us to compute Gasser-Leutwyler coefficients.
For this it seems more useful to introduce the (scale dependent) coefficients
\bqry
l_i(\mu)& = &\bar{l}_i +  \ln \left(\frac{M_{\pi}^2}{\mu^2}\right)\,=\, \ln \frac{\Lambda_i^2}{\mu^2} \lbl{Deflogl2}\,,\\[2ex]
\xi_i(\mu)& = &\bar{\xi}_i +  \ln \left(\frac{M_{\pi}^2}{\mu^2}\right)\,=\, \ln \frac{\Xi_i^2}{\mu^2} \lbl{Deflogxi2}\,,
\eqry
which are, up to irrelevant constants, the standard renormalized Gasser-Leutwyler coefficients defined in \cite{Gasser:1983yg}.\footnote{The constants are $\gamma_i/(32\pi^2)$. The $\gamma_i$ are given in eq.\ (9.6) of Ref.\ \cite{Gasser:1983yg}.} 
In terms of these coefficients the chiral logarithms are explicit and the results for the scattering lengths can be written in a more compact form. 

For simplicity let us first assume that we are interested in analyzing data at one fixed lattice spacing. In this case various terms which depend only on powers of the lattice spacing but not on the pion mass can be combined to single unknown parameters. This reduces the number of fit parameters. Explicitly we can write\footnote{The results for $a_1^1$ are summarized in appendix \ref{appC}.}
\bqry
 a_0^0&=&\phantom{-}\frac{7M_{\pi}^2}{32\pi f^2}\left(\kappa_{01}-\frac{M_{\pi}^2}{32\pi^2f^2}\left\{ 5  \ln\frac{M_{\pi}^2}{\mu^2}-\frac{40}{21}l^{\rm I=0}_{\pi\pi}\right\}
 +\frac{2c_2a^2}{32\pi^2f^2}\left\{\frac{90}{7}  \ln\frac{M_{\pi}^2}{\mu^2}\right\} \right)\nonumber\\
 &&-\frac{5\cdot2c_2a^2}{32\pi f^2}\left(\kappa_{02}+\frac{2c_2a^2}{32\pi^2f^2}\left\{11 \ln\frac{M_{\pi}^2}{\mu^2}\right\}\right)\,,\lbl{a0fit}\\[2ex]
 a_0^2&=&-\frac{M_{\pi}^2}{16\pi
 f^2}\left(\kappa_{21} + \frac{M_{\pi}^2}{16\pi^2f^2}\left\{\frac{7}{2} \ln\frac{M_{\pi}^2}{\mu^2} - \frac{4}{3}l^{\rm I=2}_{\pi\pi}\right\}
+\frac{2c_2a^2}{16\pi^2f^2}\left\{3\ln\frac{M_{\pi}^2}{\mu^2} \right\} \right)
 \nonumber\\
 &&- \frac{2c_2a^2}{16\pi f^2}\left(\kappa_{22} + \frac{2c_2a^2}{16\pi^2 f^2}\left\{\frac{11}{2}\ln\frac{M_{\pi}^2}{\mu^2}\right\} \right)\,.\lbl{a2fit}
\eqry 
The new parameters $\kappa_{{\rm I}j}$ comprise the analytic terms through O($a^2$), hence these are constants for fixed $a$, except for the fact that they are scale dependent since they contain the parameters $\xi_i(\mu)$. In the limit $a\rightarrow 0$ they assume the value $\kappa_{{\rm I}j}=1$. As a short hand notation we introduced
\bqry
l^{\rm I=0}_{\pi\pi} & =& l_1 + 2l_2 -\frac{3}{8} l_3 + \frac{21}{8}\,,\qquad
l^{\rm I=2}_{\pi\pi} \, =\, l_1 + 2l_2 -\frac{3}{8} l_3 + \frac{3}{8}\,,\lbl{LECcomb}
\eqry
for the combinations of LECs entering the scattering lengths. We emphasize again that we have expressed our results in terms of the decay constant in the chiral limit, and not in terms of $f_{\pi}$. If the latter had been used the LEC $l_4$ \cite{Gasser:1983yg} would also appear in eq.\ \pref{LECcomb}.  

Each of the expressions eqs.\ \pref{a0fit} and \pref{a2fit} contain five unknown parameters: the continuum parameters $f$ and $l_{\pi\pi}^{\rm I}$ as well as $\kappa_{{\rm I}1}, \kappa_{{\rm I}2},c_2$, i.e.\ three more than the continuum result. This is already quite large, taking into account that one usually has data for a few pion masses only. In O($a$) improved theories we have $\kappa_{{\rm I}1}=1+{\rm O}(a^2)$, 
so in this case one may try to ignore the higher order corrections and set $\kappa_{{\rm I}1}=1$.

We remark that the ratio $a_0^{ I}/M_{\pi}^2$ has the functional form
\bqry
\lbl{loopadivmpi}
\frac{a_0^{ I}}{M_{\pi}^2} & =& \frac{A_{00}}{M_{\pi}^2} + A_{10} + A_{20}M_{\pi}^2 + A_{30}M_{\pi}^2\ln M_{\pi}^2 + A_{40}\ln M_{\pi}^2+ \tilde{A}_{40}\frac{\ln M_{\pi}^2}{ M_{\pi}^2}\,.
\eqry
The first two terms on the right hand side correspond to the tree level result in eq.\ \pref{treeadivmpi}.  The constants $A_{00} - A_{40}$ represent the five independent fit parameters, while $\tilde{A}_{40}$ is not independent. The first three terms in \pref{loopadivmpi} have already been used in analyzing numerical lattice data \cite{Yamazaki:2003za}, but the data  could not be fitted well.  It might be interesting to repeat the analysis with the full result in \pref{loopadivmpi}, even though the data was obtained for heavy pion masses between 500 MeV and 1.1 GeV, and ChPT is not expected to be applicable.

If one wants to simultaneously analyze data for various lattice spacings one has to keep in mind that the coefficients $\kappa_{{\rm I}j}$ are no longer constants but functions of the lattice spacing, 
\bqry
\lbl{kappa}
\kappa_{{\rm I}j} & = & 1 + \kappa_{{\rm I}j}^{(1)} a + \kappa_{{\rm I}j}^{(2)}a^2\,.
\eqry
This increases the number of free parameters in each formula from five to seven for unimproved theories. In O($a$) improved theories we have $\kappa_{{\rm I}1}^{(1)}=0$ and the number of free parameters is increased by only one.
Note that for unimproved theories the O($a)$ corrections are not independent,
\bqry
\lbl{kapparel}
7\kappa_{01}^{(1)} & =& -5 \kappa_{21}^{(1)},\qquad  \kappa_{02}^{(1)} \,=\,  \kappa_{22}^{(1)}\,.
\eqry
Hence, if data for both scattering lengths are available the parameters in a simultaneous fit are related.  

The fit formulae for the GSM regime are obtained from eqs.\ \pref{a0fit}, \pref{a2fit} by dropping the appropriate higher order terms, leading to
\bqry
 a_0^0&=&\phantom{-}\frac{7M_{\pi}^2}{32\pi f^2}\left(\kappa_{01}-\frac{M_{\pi}^2}{32\pi^2f^2}\left\{ 5  \ln\frac{M_{\pi}^2}{\mu^2}-\frac{40}{21}l^{\rm I=0}_{\pi\pi}\right\} \right) -\frac{5\cdot2c_2a^2}{32\pi f^2}\,,\lbl{a0fitGSM}\\[2ex]
 a_0^2&=&-\frac{M_{\pi}^2}{16\pi
 f^2}\left(\kappa_{21} + \frac{M_{\pi}^2}{16\pi^2f^2}\left\{\frac{7}{2} \ln\frac{M_{\pi}^2}{\mu^2} - \frac{4}{3}l^{\rm I=2}_{\pi\pi}\right\}
 \right) - \frac{2c_2a^2}{16\pi f^2}\,.\lbl{a2fitGSM}
\eqry 
Here $\kappa_{{\rm I}j}  = 1 + \kappa_{{\rm I}j}^{(1)} a$. The free fit parameters are $f,c_2$ and $\kappa_{{\rm I}j}^{(1)}$, and the latter vanish in O($a$) improved theories. These formulae lead to the general form 
\bqry
\lbl{loopadivmpiGSM}
\frac{a_0^{I}}{M_{\pi}^2} & =& \frac{A_{00}}{M_{\pi}^2} + A_{10} + A_{20}M_{\pi}^2 + A_{30}M_{\pi}^2\ln M_{\pi}^2,
\eqry
for the ratio $a_0^{\rm I}/M_{\pi}^2$, in contrast to \pref{loopadivmpi}. 

\section{Concluding remarks}
\label{sect:conclusion}

Present day lattice simulations are still done with quark masses much heavier than in nature. Therefore, a chiral extrapolation to the physical point is still necessary, and that is where the predictions of ChPT enter the analysis of numerical lattice data. However, the results of continuum ChPT get modified at non-zero lattice spacing.  
For pion scattering with Wilson fermions we essentially find two modifications. First, the  $I=0,2$ scattering lengths do not vanish in the chiral limit, but rather assume a nonzero value of order $a^2$. Second, additional chiral logarithms proportional to $a^2$ appear in the one-loop results for these quantities.  
Ignoring these modifications and using the results of continuum ChPT is potentially dangerous, 
depending on the size of these extra contributions. One either introduces a systematic error in the chiral extrapolation or the data cannot be fitted at all with the continuum  results.  

Related problems may arise in the determination of the Gasser-Leutwyler coefficients associated with pion scattering. 
Experience shared by many lattice groups is that the lattice data does not show the characteristic curvature due to the continuum chiral logarithms. A possible explanation is the presence of additional chiral logarithms proportional to the lattice spacing. These can conspire with the continuum chiral logarithms such that the overall curvature 
of the data is diminished. A correct and precise determination of the Gasser-Leutwyler coefficients using the continuum results is very unlikely in that case. Using the expression derived here should help in that respect.

The formulae we presented in this paper involve the parameter $c_2$, which also determines the phase diagram of the theory. In particular the $I=2$ scattering length may provide a handle to obtain an estimate of $c_2$. At least the sign of $c_2$ should be easily accessible, and this is what matters for the phase diagram.

We conclude with a remark on finite volume corrections. The numerical calculation of phase shifts and scattering lengths is usually done by employing the so-called L\"uscher formula \cite{Luscher:1986pf,Luscher:1990ux} in order to circumvent the Maiani-Testa no-go theorem \cite{Maiani:1990ca}. This formula relates the two-pion energy eigenvalues in finite volume to the infinite volume scattering length of the two-pion scattering process.

In addition to the power law finite volume dependence of the the two-pion energy eigenvalues, which one exploits to extract the infinite volume scattering lengths, there are exponentially suppressed finite volume corrections. These have been studied in Ref.\ \cite{Bedaque:2006yi} and it is straightforward to include these in  our results. However, these corrections are expected to be very small on typical lattice sizes, much smaller than the corrections due to the non-zero lattice spacing. For example, for a pion mass of approximately 300 MeV and a finite volume with $L\simeq 2.5$ fm the finite volume correction to the $I=2$ scattering length is about 1 to 2\% \cite{Bedaque:2006yi}.\footnote{For this estimate we made the crude approximation $k\cot \delta \approx 1/a_0^2$.} On the other hand, taking $|2c_2 a^2| \approx (185 \, {\rm MeV})^2$ at $a\approx 0.086$ fm found by the ETM collaboration at a charged pion mass of about 300 MeV \cite{Michael:2007vn,Urbach:2007rt}, we obtain the rough estimate of about 35\% for the O($a^2$) corrections in this channel. Although the numerical value for  $2c_2 a^2$ has a large error bar and can easily be a factor of 2 or 4 smaller, these numbers indicate that the finite volume corrections are most likely much smaller than the lattice spacing corrections.  

\section*{Acknowledgments}

This work is supported in part by the Grants-in-Aid for
Scientific Research from the Japanese Ministry of Education,
Culture, Sports, Science and Technology
(No.\ 20340047) and by the Deutsche Forschungsgemeinschaft (SFB/TR 09).
B. B.\  acknowledges financial support from the {\em Cusanuswerk}.

\begin{appendix}

\section{Higher order terms in the chiral Lagrangian}
\label{appA}
The terms required for one loop calculations in the LCE regime are shown in \pref{PcountingLCE}. Among them are the NLO terms of continuum ChPT, which are given in Ref.\ \cite{Gasser:1983yg}. The terms of O$(p^{2}a,ma,a^2)$ can be found in Refs.\ \cite{Rupak:2002sm,Bar:2003mh}. So far unknown are the contributions of order $p^2a^2, ma^2,a^3,a^4$. 
However, these missing terms are easily constructed with the spurion fields introduced in \cite{Bar:2003mh}. 

There are essentially two independent spurion fields, $M$ and $A$. These stem from the two sources of explicit chiral symmetry breaking, the quark mass and the lattice spacing. Under chiral symmetry transformations these fields transform as
\bqry
\lbl{trafoSpurion}
M&\longrightarrow & LMR^{\dagger},\qquad A\,\longrightarrow \, LAR^{\dagger}\,.
\eqry
These fields together with the field $\Sigma$ and its derivatives are used to write down the most general chiral Lagrangian that is compatible with chiral symmetry, parity and charge conjugation. Once the terms in the chiral Lagrangian have been found the spurion fields are assigned to their physical values,
\bqry
\lbl{finalvalueSpurion}
M&\longrightarrow & {\rm diag}(m_u,m_d)\,=\, mI ,\qquad A\,\longrightarrow \, aI\,,
\eqry
where $I$ denotes the two dimensional unit matrix (recall that we ignore isospin violation and assume $m_u=m_d\equiv m$). The number of terms in the chiral Lagrangian can be reduced further by making use of the Cayley-Hamilton theorem.

Proceeding along these lines we  find the following independent terms of O($p^2a^2,ma^2$):
\bqry
\lbl{masqr}
{\cal L}_{[p^2a^2]} &=&\phantom{+}  a_1 a^2 \langle \partial_{\mu}\Sigma  \partial_{\mu}\Sigma^{\dagger} \rangle + a_2 a^2 \langle \partial_{\mu}\Sigma  \partial_{\mu}\Sigma^{\dagger} \rangle \langle \Sigma+\Sigma^{\dagger}\rangle^2\nonumber\\
& & \!\!+\, a_3 a^2 \langle \partial_{\mu}(\Sigma +\Sigma^{\dagger}) \rangle\langle \partial_{\mu}(\Sigma +\Sigma^{\dagger}) \rangle\,,\\
{\cal L}_{[ma^2]} &= & b_1 ma^2 \langle \Sigma + \Sigma^{\dagger}\rangle + b_2 ma^2 \langle \Sigma + \Sigma^{\dagger}\rangle^3\,.
\eqry
The coefficients $a_j,b_j$ are undetermined LECs. 

Since the spurion fields $M$ and $A$ transform identically under all symmetries we
trivially obtain the O($a^3$) terms by replacing $M$ with $A$ in the O($ma^2$) terms. This leads to 
\bqry
\lbl{aqubic}
{\cal L}_{[a^3]} &= & d_1 a^3 \langle \Sigma + \Sigma^{\dagger}\rangle + d_2 a^3 \langle \Sigma + \Sigma^{\dagger}\rangle^3\,.
\eqry
Finally, for the O($a^4$) terms we obtain
\bqry
\lbl{aquartic}
{\cal L}_{[a^4]} &= & e_1 a^4 \langle \Sigma + \Sigma^{\dagger}\rangle^2 + e_2 a^4 \langle \Sigma + \Sigma^{\dagger}\rangle^4\,.
\eqry
Here we have dropped the constant term $e_0 a^4$, which is also admitted by the symmetries but does not contribute to the pion mass and scattering amplitude. 

Note that we have obtained the O($a^3,a^4$) terms although the Symanzik effective action has not been constructed to these orders. However, any spurion field at these orders has to transform as an appropriate  tensor product of the spurion field $A$ and therefore gives rise to the terms in \pref{aqubic}, \pref{aquartic} only.

\section{The scattering amplitude to one loop}
\label{appB}
The form of the one-loop result for the scattering amplitude is defined in eq.\ \pref{scatamponeloop}. The functions $B(s,t,u),C(s,t,u)$, introduced in eqs.\ \pref{Bfunction}, \pref{Cfunction},  contain the one-loop corrections. The parts $B_{\rm cont},C_{\rm cont}$ are the contributions from continuum ChPT and read 
\bqry
B_{\rm cont} & =& \frac{1}{96\pi^2f^4}\bigg\{ 3({s}^{2}-{M}_0^4)
F \left( s \right) \nonumber\\
 &&\phantom{\frac{1}{96\pi^2f^2}\bigg\{} +\{ t \left( t-u \right) -2{M}_0^2 \left( t-2u \right) -2{M}_0^4 \}F (t)\nonumber\\
 &&\phantom{\frac{1}{96\pi^2f^2}\bigg\{}+\{ u \left( u-t \right) -2{M}_0^2 \left( u-2t \right) -2{M}_0^4) \}F \left( u \right) \bigg\}\,,\\
C_{\rm cont} & =& \frac{1}{96\pi^2f^4}\bigg\{2\, \left( {\bar{l}_1}+2/3 \right)  ( s-2\,{M}_0^{2} ) ^{2}+ \left( {\bar{l}_2}+7/6 \right)  \left( {s}^{2}+ ( t-u \right) ^{2}
 ) +{M}_0^{4}\bigg\}\nonumber
 \eqry
The constants $\bar{l}_i$ are defined in eq.\ \pref{Deflogl}, $M_0^2$ is the tree-level pion mass of eq.\  \pref{Mpitree},  and the function $F(x)$ is given by
\bqry
\lbl{functionF}
F(x) &=& -\sigma\left(\ln\frac{1+\sigma}{1-\sigma} - i\pi\right),\quad \sigma\,=\,\sqrt{1-\frac{4 M_0^2}{x}}\,.
\eqry
These results agree with the ones in \cite{Gasser:1983kx}. The only difference is the use of the function $\bar{J}(x)$ in \cite{Gasser:1983kx}, which is related to our $F(x)$ by
\bqry
\lbl{relJF}
16\pi^2 \bar{J}(x) & = & F(x) + 2.
\eqry
This leads to some differences between our functions $B_{\rm cont}, C_{\rm cont}$ and the ones in \cite{Gasser:1983kx}. The result for the scattering amplitude $A(s,t,u)$, however, is the same.

For the O($a^2$) contributions we find
\bqry
B_{a^2}(s,t,u) & = &\frac{1}{96\pi^2f^4}\bigg\{(42c_2a^2+18M_0^2-24s )
F \left( s \right) \nonumber\\
&&\phantom{\frac{1}{96\pi^2f^4}\big\{}+(12c_2a^2-12 M_0^2+6t) F(t) \nonumber\\
&&\phantom{\frac{1}{96\pi^2f^4}\big\{}+(12c_2a^2-12M_0^2+6u) F(u) \bigg\}\,,\lbl{Basqr}\\
C_{a^2}(s,t,u) & = & \frac{1}{96\pi^2f^4}\bigg\{ -30s ( \bar{\xi}_4+1) +3M_0^2 ( 11\,\bar{\xi}_5+6)\nonumber\\
&&\phantom{\frac{1}{96\pi^2f^4}\bigg\{}+66c_{2}a^{2}
(\bar{\xi}_6+1 ) \bigg\} + k_5\frac{W_0a}{f^4}.\lbl{Casqr}
\eqry
The constants $\bar{\xi}_i= \bar{\xi}_i(M_0^2)$ are defined in eq.\ \pref{Deflogxi} and $k_5$ is the LEC associated with the O($a^3$) terms (it is a combination of the $c_j$ in \pref{aqubic}). As claimed in section \ref{ssect:scatteringoneloop}, these functions assume a finite value for $a\rightarrow 0$. Hence the scattering amplitude reduces to the continuum result in this limit (recall the additional factor $2c_2a^2$ in front of $B_{a^2},C_{a^2}$ in eqs.\ \pref{Bfunction}, \pref{Cfunction}). 

\section{The $\mathbf{I=1}$ scattering length}
\label{appC}

With the result for the scattering amplitude of the previous section we can compute other quantities like the slope parameters $b_l^I$ or the phase shifts $\delta_l^I$. Here we only quote the one-loop result for the scattering length $a_1^1$ in the isospin one channel, because the corrections due to a non-zero lattice spacing are fewer than in the other two isospin channels.  

Performing the partial wave expansion for the $I=1$ case we straightforwardly obtain
\bqry
 a_1^1&=&\phantom{+}\frac{M_0^2}{24\pi
 f^2}\left(1-\frac{M_0^2}{12\pi^2f^2}\left\{\bar{l}_1(M_0^2)-\bar{l}_2(M_0^2)+\frac{65}{48}\right\}-\frac{2c_2a^2}{16\pi^2f^2}\left\{5\bar{\xi}_4(M_0^2)-\frac{35}{6}\right\}\right)\nonumber\\
 &&+\frac{2c_2a^2}{24\pi
 f^2}\left(\frac{2c_2a^2}{16\pi^2f^2}\left\{\frac{5}{12}\right\}\right)\lbl{resulta1}
\eqry
for the scattering length. We recover the continuum result in Ref.\ \cite{Gasser:1983yg} for a vanishing lattice spacing. 

Note that the number of lattice spacing corrections is reduced compared to the results \pref{resulta0}, \pref{resulta2} for the other two isospin channels. There are no corrections proportional to $a^2$ and $a^3$ in the result for $a_1^1$. Also the $a^4\ln \mpit$ term is missing. The absence of these terms is a direct consequence of taking the difference $A(t,s,u)-A(u,t,s)$ for the $I=1$ scattering amplitude.
However, analytic corrections of O($a^3)$ enter when we replace $\mpit$ by $M_{\pi}^2$ and express the scattering length as a function of the pion mass: 
\bqry
 a_1^1&=&\phantom{+}\frac{M_{\pi}^2}{24\pi
 f^2}\left(1 - \frac{M_{\pi}^2}{12\pi^2f^2}\left\{\bar{l}_1(M_{\pi}^2)-\bar{l}_2(M_{\pi}^2)-\frac{3}{8}\bar{l}_3(M_{\pi}^2)+\frac{65}{48}\right\}\right.\nonumber\\
 & &\phantom{\frac{M_{\pi}^2}{24\pi
 f^2}\bigg(}\left.\,\,-k_1\frac{W_0 a}{f^2}-\frac{2c_2a^2}{16\pi^2f^2}\left\{-\frac{5}{2}\bar{\xi}_3(M_{\pi}^2)+ 5\bar{\xi}_4(M_{\pi}^2)-\frac{35}{6}\right\}\right)\nonumber\\
 &&-\frac{2c_2a^2}{24\pi
 f^2}\left(k_3\frac{W_0a}{f^2} +k_4\frac{2c_2a^2}{f^2}-\frac{2c_2a^2}{16\pi^2f^2}\left\{\frac{5}{12}\right\}\right)\lbl{resulta1two}
\eqry
For completeness we also give the analogue of the fit formulae in section \ref{ssect:Disc}. Replacing the 
$\bar{l}_i$ and $\bar{\xi}_3$ using eqs.\ \pref{Deflogl}, \pref{Deflogxi} we can write
\bqry
a_1^1&=&\phantom{-}\frac{M_{\pi}^2}{24\pi
 f^2}\left(\kappa_{11} - \frac{M_{\pi}^2}{24\pi^2f^2}\left\{\frac{3}{4} \ln\frac{M_{\pi}^2}{\mu^2} + 2l^{\rm I=1}_{\pi\pi}\right\}
+\frac{2c_2a^2}{24\pi^2f^2}\left\{\frac{15}{4}\ln\frac{M_{\pi}^2}{\mu^2} \right\} \right)
 \nonumber\\
 &&-\frac{2c_2a^2}{24\pi f^2}\,\kappa_{12}\,,\lbl{a1fit}
 \eqry
 where we introduced the combination
 \bqry
l^{\rm I=1}_{\pi\pi} & =& l_1 -l_2 -\frac{3}{8} l_3 + \frac{65}{48}\,.
\eqry
The coefficients $\kappa_{1j}$ comprise the analytic terms through O($a^2$). They are of the form
\bqry
\lbl{kappaI1}
\kappa_{11} & = & 1 + \kappa_{11}^{(1)} a+ \kappa_{11}^{(2)} a^2\,,\qquad \kappa_{12}\,=\, \kappa_{12}^{(1)} a + \kappa_{12}^{(2)}a^2\,.
\eqry
These coefficients are constants for a fixed lattice spacing. The coefficient $\kappa_{11}^{(1)}$ is related to the corresponding coefficients for the other two isospin channels:
\bqry
\lbl{kapparel2}
7\kappa_{01}^{(1)} & =& -5 \kappa_{21}^{(1)}\,=\,- 5\kappa_{11}^{(1)}.
\eqry
In the O($a$) improved theory all these coefficients vanish.
 
There is no relation for $\kappa_{12}^{(1)}$ since it does not involve the low-energy constant $k_5$. 
Note that $\kappa_{12}$ vanishes for $a\rightarrow0$, in contrast to the other two isospin channels. 

Finally, in order to obtain the result for the GSM regime one has to drop the $a^2M^2_{\pi}\ln M^2_{\pi}$ term and sets $\kappa_{11}=1$ and $\kappa_{12}=0$. The resulting expression is simply the one-loop result of continuum ChPT.

\end{appendix}

\end{document}